\documentclass[onecolumn,showpacs,preprint,ams,floats]{revtex4}
\usepackage{amsmath}
\usepackage{graphics, graphicx}
\usepackage{verbatim}

\input{tcilatex}

\begin{document}

\title{Permutation-invariant monotones for multipartite entanglement
characterization}
\author{Xi-Jun Ren}
\email{renxijun@mail.ustc.edu.cn}
\affiliation{Key Laboratory of Quantum Information, University of Science and Technology
of China, Chinese Academy of Sciences, Hefei, Anhui 230026, China}
\author{Wei Jiang}
\affiliation{Key Laboratory of Quantum Information, University of Science and Technology
of China, Chinese Academy of Sciences, Hefei, Anhui 230026, China}
\author{Xingxiang Zhou}
\email{xizhou@yahoo.com}
\affiliation{Key Laboratory of Quantum Information, University of Science and Technology
of China, Chinese Academy of Sciences, Hefei, Anhui 230026, China}
\author{Zheng-Wei Zhou}
\email{zwzhou@ustc.edu.cn}
\affiliation{Key Laboratory of Quantum Information, University of Science and Technology
of China, Chinese Academy of Sciences, Hefei, Anhui 230026, China}
\author{Guang-Can Guo}
\affiliation{Key Laboratory of Quantum Information, University of Science and Technology
of China, Chinese Academy of Sciences, Hefei, Anhui 230026, China}

\begin{abstract}
In this work we consider the permutational properties of multipartite
entanglement monotones. Based on the fact that genuine multipartite
entanglement is a property of the entire multi-qubit system, we argue that
ideal definitions for its characterizing quantities must be
permutation-invariant. Using this criterion, we examine the three 4-qubit
entanglement monotones introduced by Osterloh and Siewert [Phys. Rev. A. 
\textbf{72}, 012337]. By expressing them in terms of quantities whose
permutational properties can be easily derived, we find that one of these
monotones is not permutation-invariant. We propose a permutation-invariant
entanglement monotone to replace it, and show that our new monotone properly
measures the genuine 4-qubit entanglement in 4-qubit cluster-class states.
Our results provide some useful insights in understanding multipartite
entanglement.
\end{abstract}

\maketitle

Quantification of quantum entanglement is one of the most important problems
in quantum information theory \cite{Nielsen}. For pure states of bipartite
systems, a good entanglement measure is entropy of entanglement \cite%
{Bennett}. It measures the entanglement content in a pure bipartite state
which can be asymptotically concentrated into the standard form of Bell
states, using local operations and classical communications (LOCC). For pure
states of multi-qubit systems, the problem of defining good entanglement
measures becomes complicated because the qubits in a generic multi-qubit
state can be entangled in different ways under asymptotic LOCC \cite%
{Bennett1}.

Following recent studies on the structures of genuine multipartite
entanglement \cite{Dur, Verstraete1, Miyake}, researchers have proposed many
quantitative measures for different classes of multipartite entanglement 
\cite{Jaeger, Lamata, Wong, Meyer, Heydari}. A particular interesting
proposal is Osterloh and Siewert \cite{Osterloh,Osterloh1}'s N-qubit
entanglement monotones (for an N-qubit system) constructed with antilinear
operators. They have the appealing property of yielding zero for all
possible product states, and are therefore considered good measures for
genuine N-qubit entanglement \cite{Osterloh1, Cai}.

Despite the recent progress in the study of multipartite entanglement, the
important issue of permutational properties of its quantitative measures has
not received sufficient attention. For an N-qubit system, genuine N-qubit
entanglement is an overal property of the entire system and should not rely
on the labeling of the qubits. In other words, proper definitions for
quantities characterizing genuine multipartite entanglement should be
permutation-invariant. This is a natural requirement and is obviously
satisfied by the bipartite entropy of entanglement. However, the
permutational properties of most existing definitions of multipartite
entanglement measures are rather hard to investigate because of their
mathematical complexity. As an attempt to study the permutational properties
of multipartite entanglement measures, we analyze Osterloh and Siewert's
4-qubit entanglement monotones \cite{Osterloh,Osterloh1} and show that one
of them is not permutation-invariant. In place of it, we propose a monotone
which does preserve its value under qubit permutation. We show that our
monotone properly describes the genuine 4-qubit entanglement in 4-qubit
cluster-class states and thus is a good alternative to the
non-permutation-invariant quantity in Osterloh and Siewert's monotones.

We start by briefly introducing Osterloh and Siewert's 4-qubit entanglement
monotones for a generic 4-qubit state 
\begin{equation}
\left\vert \Psi \right\rangle =\sum\limits_{i,j,k,l=0}^{1}A_{ijkl}\left\vert
i\right\rangle \otimes \left\vert j\right\rangle \otimes \left\vert
k\right\rangle \otimes \left\vert l\right\rangle .  \label{state}
\end{equation}%
The three entanglement monotones, named filters in \cite{Osterloh}, are
constructed with antilinear operators as follows: 
\begin{equation}
\mathcal{F}_{1}=(\sigma _{\mu }\sigma _{\nu }\sigma _{y}\sigma _{y})\bullet
(\sigma ^{\mu }\sigma _{y}\sigma _{\lambda }\sigma _{y})\bullet (\sigma
_{y}\sigma ^{\nu }\sigma ^{\lambda }\sigma _{y}),  \label{filter1}
\end{equation}%
\begin{equation}
\mathcal{F}_{2}=(\sigma _{\mu }\sigma _{\nu }\sigma _{y}\sigma _{y})\bullet
(\sigma ^{\mu }\sigma _{y}\sigma _{\lambda }\sigma _{y})\bullet (\sigma
_{y}\sigma ^{\nu }\sigma _{\tau }\sigma _{y})\bullet (\sigma _{y}\sigma
_{y}\sigma ^{\lambda }\sigma ^{\tau }),  \label{filter2}
\end{equation}%
\begin{eqnarray}
\mathcal{F}_{3} &=&\frac{1}{2}(\sigma _{\mu }\sigma _{\nu }\sigma _{y}\sigma
_{y})\bullet (\sigma ^{\mu }\sigma ^{\nu }\sigma _{y}\sigma _{y})\bullet
(\sigma _{\rho }\sigma _{y}\sigma _{\tau }\sigma _{y})  \notag \\
&&\bullet (\sigma ^{\rho }\sigma _{y}\sigma ^{\tau }\sigma _{y})\bullet
(\sigma _{y}\sigma _{\lambda }\sigma _{\xi }\sigma _{y})\bullet (\sigma
_{y}\sigma ^{\lambda }\sigma ^{\xi }\sigma _{y}),  \label{filter3}
\end{eqnarray}%
where the same index is contracted with the metric $g^{\mu ,\nu
}=diag\{-1,1,0,1\}$. Using these operators, we can obtain three SLOCC
invariants and entanglement monotones \cite{Bennett1,Vidal}: $\left\vert 
\mathcal{F}_{i}\right\vert =\left\vert \left\langle \left\langle \mathcal{F}%
_{i}\right\rangle \right\rangle _{C}\right\vert $, $i=1,2,3$. Here, a
complex conjugation is carried out before the expectation is taken. By
virtue of the special contraction rules, these monotones have zero values
for all possible product states and therefore characterize genuine 4-qubit
entanglement.

As shown in \cite{Osterloh}, $\left\vert \mathcal{F}_{i}\right\vert $'s can
be used to classify different 4-qubit entangled states. Whether they qualify
as proper genuine 4-qubit entanglement measures is an open question, however
their permutational properties provide an important criterion in resolving
this matter. Unfortunately, it is difficult to see $\left\vert \mathcal{F}
_{i}\right\vert $'s permutational properties directly from their antilinear
operator definitions given above. To attack this problem, we adopt an
indirect strategy by expressing $\left\vert \mathcal{F}_{i}\right\vert $'s
in terms of a set of $SL^{4}(2)$ algebraic invariants introduced by Luque
and Thibon \cite{Luque} whose permutational properties can be easily deduced.

Considering the fact that many SLOCC invariants can be constructed with the
antisymmetric tensor $\epsilon $ \cite{Leifer}, we first write $\left\vert 
\mathcal{F}_{i}\right\vert $'s using $\epsilon $ and $A_{ijkl}$, the
coefficients of a generic 4-qubit pure state in Eq. (\ref{state}). This can
be done by expressing $\sigma $'s, the so-called \textquotedblleft
combs\textquotedblright\ in the definitions of $\left\vert \mathcal{F}
_{i}\right\vert $'s, in terms of $\epsilon $. The first comb operator, $%
\sigma _{y},$ can be directly replaced with antisymmetric tensor $\epsilon $
because of the relation $\sigma _{y}=-i\epsilon $. The second comb operator, 
$\sigma _{\mu }\bullet \sigma ^{\mu },$ where the upper and lower indexes
are contracted with the metric $g^{\mu ,\nu }=diag\{-1,1,0,1\}$, can also be
expressed with $\epsilon $. For example, if $\{i_{1},i_{2},i_{3},i_{4}\}$ is
contracted under the rules of $\sigma _{\mu }\bullet \sigma ^{\mu }$, then
we have $(\sigma _{\mu })_{i_{1}i_{2}}g_{\mu \nu }(\sigma _{\nu
})_{i_{3}i_{4}}$ and with direct calculations we can check that this is
equal to $\epsilon _{i_{1}i_{3}}\epsilon _{i_{4}i_{2}}+\epsilon
_{i_{1}i_{4}}\epsilon _{i_{3}i_{2}}$. Using these relations, we obtain 
\begin{widetext}
\begin{eqnarray}
\left\vert \mathcal{F}_{1}\right\vert
&=&|4A_{i_{1}j_{1}k_{1}l_{1}}A_{i_{2}j_{2}k_{2}l_{2}}A_{i_{3}j_{3}k_{3}l_{3}}A_{i_{4}j_{4}k_{4}l_{4}}A_{i_{5}j_{5}k_{5}l_{5}}A_{i_{6}j_{6}k_{6}l_{6}}
\nonumber \\
&&\times (\epsilon _{i_{1}i_{3}}\epsilon _{i_{2}i_{4}}+\epsilon
_{i_{1}i_{4}}\epsilon _{i_{2}i_{3}})\epsilon _{i_{5}i_{6}}\epsilon
_{j_{1}j_{5}}\epsilon _{j_{3}j_{4}}\epsilon _{j_{2}j_{6}}\epsilon
_{k_{1}k_{2}}\epsilon _{k_{3}k_{5}}\epsilon _{k_{4}k_{6}}\epsilon
_{l_{1}l_{2}}\epsilon _{l_{3}l_{4}}\epsilon _{l_{5}l_{6}}|,  \label{f1}
\\
%\end{eqnarray}%
%\begin{eqnarray}
\left\vert \mathcal{F}_{2}\right\vert
&=&|8A_{i_{1}j_{1}k_{1}l_{1}}A_{i_{2}j_{2}k_{2}l_{2}}A_{i_{3}j_{3}k_{3}l_{3}}A_{i_{4}j_{4}k_{4}l_{4}}A_{i_{5}j_{5}k_{5}l_{5}}A_{i_{6}j_{6}k_{6}l_{6}}A_{i_{7}j_{7}k_{7}l_{7}}A_{i_{8}j_{8}k_{8}l_{8}}
\nonumber \\
&&\times (\epsilon _{i_{1}i_{3}}\epsilon _{i_{2}i_{4}}+\epsilon
_{i_{1}i_{4}}\epsilon _{i_{2}i_{3}})\epsilon _{i_{5}i_{6}}\epsilon
_{i_{7}i_{8}}\epsilon _{j_{1}j_{5}}\epsilon _{j_{3}j_{4}}\epsilon
_{j_{2}j_{6}}\epsilon _{j_{7}j_{8}}\epsilon _{k_{1}k_{2}}\epsilon
_{k_{3}k_{7}}\epsilon _{k_{5}k_{6}}\epsilon _{k_{4}k_{8}}\epsilon
_{l_{1}l_{2}}\epsilon _{l_{3}l_{4}}\epsilon _{l_{5}l_{7}}\epsilon
_{l_{6}l_{8}}|,  \label{f2}
\\
%\end{eqnarray}%
%\begin{eqnarray}
\left\vert \mathcal{F}_{3}\right\vert
&=&|4A_{i_{1}j_{1}k_{1}l_{1}}A_{i_{2}j_{2}k_{2}l_{2}}A_{i_{3}j_{3}k_{3}l_{3}}A_{i_{4}j_{4}k_{4}l_{4}}(\epsilon _{i_{1}i_{3}}\epsilon _{i_{2}i_{4}}+\epsilon _{i_{1}i_{4}}\epsilon _{i_{2}i_{3}})\epsilon _{j_{1}j_{3}}\epsilon _{j_{2}j_{4}}\epsilon _{k_{1}k_{2}}\epsilon _{k_{3}k_{4}}\epsilon _{l_{1}l_{2}}\epsilon _{l_{3}l_{4}}
\nonumber \\
&&\times
A_{i_{5}j_{5}k_{5}l_{5}}A_{i_{6}j_{6}k_{6}l_{6}}A_{i_{7}j_{7}k_{7}l_{7}}A_{i_{8}j_{8}k_{8}l_{8}}(\epsilon _{i_{5}i_{7}}\epsilon _{i_{6}i_{8}}+\epsilon _{i_{5}i_{8}}\epsilon _{i_{6}i_{7}})\epsilon _{j_{5}j_{6}}\epsilon _{j_{7}j_{8}}\epsilon _{k_{5}k_{7}}\epsilon _{k_{6}k_{8}}\epsilon _{l_{5}l_{6}}\epsilon _{l_{7}l_{8}}
\nonumber \\
&&\times
A_{i_{9}j_{9}k_{9}l_{9}}A_{i_{10}j_{10}k_{10}l_{10}}A_{i_{11}j_{11}k_{11}l_{11}}A_{i_{12}j_{12}k_{12}l_{12}}\epsilon _{i_{9}i_{10}}\epsilon _{i_{11}i_{12}}(\epsilon _{j_{9}j_{11}}\epsilon _{j_{10}j_{12}}+\epsilon _{j_{9}j_{12}}\epsilon _{j_{10}j_{11}})\epsilon _{k_{5}k_{7}}\epsilon _{k_{6}k_{8}}
\nonumber \\
&&\times \epsilon _{l_{9}l_{10}}\epsilon _{l_{11}l_{12}}|.  \label{f3}
\end{eqnarray}%
\end{widetext}

In \cite{Luque}, Luque and Thibon introduced a complete and independent set
of four $SL^{4}(2)$ algebraic invariants, $\{H,L,M,D_{xt}\}$, for a 4-qubit
system. The permutational properties of these invariants are simple and will
be derived later. Our aim is to write $\left\vert \mathcal{F}_{i}\right\vert 
$'s with this set of invariants, by comparing their antisymmetric tensor
expressions. $H$, a degree-2 invariant whose each term involves only two
coefficients, can be directly written as 
\begin{equation}
H=\frac{1}{2}A_{i_{1}j_{1}k_{1}l_{1}}A_{i_{2}j_{2}k_{2}l_{2}}\epsilon
_{i_{1}i_{2}}\epsilon _{j_{1}j_{2}}\epsilon _{k_{1}k_{2}}\epsilon
_{l_{1}l_{2}}.  \label{H}
\end{equation}%
The three degree-4 invariants $L,M,N$ are given by the determinants of three
matrices: 
\begin{equation}
L=\left\vert 
\begin{array}{cccc}
a_{0} & a_{4} & a_{8} & a_{12} \\ 
a_{1} & a_{5} & a_{9} & a_{13} \\ 
a_{2} & a_{6} & a_{10} & a_{14} \\ 
a_{3} & a_{7} & a_{11} & a_{15}%
\end{array}%
\right\vert ,M=\left\vert 
\begin{array}{cccc}
a_{0} & a_{8} & a_{2} & a_{10} \\ 
a_{1} & a_{9} & a_{3} & a_{11} \\ 
a_{4} & a_{12} & a_{6} & a_{14} \\ 
a_{5} & a_{13} & a_{7} & a_{15}%
\end{array}%
\right\vert ,N=\left\vert 
\begin{array}{cccc}
a_{0} & a_{1} & a_{8} & a_{9} \\ 
a_{2} & a_{3} & a_{10} & a_{11} \\ 
a_{4} & a_{5} & a_{12} & a_{13} \\ 
a_{6} & a_{7} & a_{14} & a_{15}%
\end{array}%
\right\vert ,  \label{LMN}
\end{equation}%
where $a_{r}=A_{ijkl},r=8i+4j+2k+l$, is another notation of state
coefficients. $L,M,N$ have the relation $L+M+N=0$, thus only two of them are
linearly independent. It should be noted that $H^{2}$ cannot be expressed as
a linear combination of $\{L,M,N\},$ thus $H$ is independent of them. $%
D_{xt} $, the last in the set $\{H,L,M,D_{xt}\},$ is of degree 6. All
together, there are 6 degree-6 invariants $%
D_{xt},D_{xy},D_{xz},D_{yz},D_{yt},D_{zt}.$ These six invariants are
constructed with methods of classical invariants theory. They satisfy $%
D_{xt}=D_{yz},D_{xy}=D_{zt},D_{xz}=D_{yt}$ \cite{Luque} and can be expressed
as the determinants of three $3\times 3$ matrices:

\begin{widetext}
\begin{equation*}
D_{xt}=\left\vert
\begin{array}{ccc}
a_{0}a_{6}-a_{2}a_{4}, & a_{0}a_{7}+a_{1}a_{6}-a_{2}a_{5}-a_{3}a_{4}, &
a_{1}a_{7}-a_{3}a_{5}, \\
\begin{array}{c}
a_{0}a_{14}+a_{8}a_{6} \\
-a_{2}a_{12}-a_{4}a_{10},%
\end{array}
&
\begin{array}{c}
a_{0}a_{15}+a_{6}a_{9}+a_{1}a_{14}+a_{7}a_{8} \\
-a_{2}a_{13}-a_{4}a_{11}-a_{3}a_{12}-a_{5}a_{10},%
\end{array}
&
\begin{array}{c}
a_{1}a_{15}+a_{7}a_{9} \\
-a_{3}a_{13}-a_{5}a_{11},%
\end{array}
\\
a_{8}a_{14}-a_{10}a_{12}, &
a_{8}a_{15}+a_{9}a_{14}-a_{10}a_{13}-a_{11}a_{12}, &
a_{9}a_{15}-a_{11}a_{13}.%
\end{array}%
\right\vert ,
\end{equation*}%
\begin{equation*}
D_{xy}=\left\vert
\begin{array}{ccc}
a_{0}a_{3}-a_{1}a_{2}, & a_{0}a_{7}+a_{3}a_{4}-a_{2}a_{5}-a_{1}a_{6}, &
a_{4}a_{7}-a_{5}a_{6}, \\
\begin{array}{c}
a_{0}a_{11}+a_{3}a_{8} \\
-a_{2}a_{9}-a_{1}a_{10},%
\end{array}
&
\begin{array}{c}
a_{0}a_{15}+a_{3}a_{12}+a_{4}a_{11}+a_{7}a_{8} \\
-a_{2}a_{13}-a_{1}a_{14}-a_{6}a_{9}-a_{5}a_{10},%
\end{array}
&
\begin{array}{c}
a_{4}a_{15}+a_{7}a_{12} \\
-a_{6}a_{13}-a_{5}a_{14},%
\end{array}
\\
a_{8}a_{11}-a_{9}a_{10}, & a_{8}a_{15}+a_{11}a_{12}-a_{10}a_{13}-a_{9}a_{14},
& a_{12}a_{15}-a_{13}a_{14}.%
\end{array}%
\right\vert ,
\end{equation*}%
\begin{equation*}
D_{xz}=\left\vert
\begin{array}{ccc}
a_{0}a_{5}-a_{1}a_{4}, & a_{0}a_{7}+a_{2}a_{5}-a_{1}a_{6}-a_{3}a_{4}, &
a_{2}a_{7}-a_{3}a_{6}, \\
\begin{array}{c}
a_{0}a_{13}+a_{5}a_{8} \\
-a_{1}a_{12}-a_{4}a_{9},%
\end{array}
&
\begin{array}{c}
a_{0}a_{15}+a_{5}a_{10}+a_{2}a_{13}+a_{7}a_{8} \\
-a_{1}a_{14}-a_{4}a_{11}-a_{3}a_{12}-a_{6}a_{9},%
\end{array}
&
\begin{array}{c}
a_{2}a_{15}+a_{7}a_{10} \\
-a_{3}a_{14}-a_{6}a_{11},%
\end{array}
\\
a_{8}a_{13}-a_{9}a_{12}, & a_{8}a_{15}+a_{10}a_{13}-a_{9}a_{14}-a_{11}a_{12},
& a_{10}a_{15}-a_{11}a_{14}.%
\end{array}%
\right\vert .
\end{equation*}
\end{widetext}

$D_{xt},D_{xy}$ and $D_{xz}$ are linearly independent. They have the
following relations with \{$H,L,M,N$\} \cite{Luque}: $HL=D_{xz}-D_{xt},$ $%
HM=D_{xt}-D_{xy},$ $HN=D_{xy}-D_{xz}.$ Therefore only one of \{$%
D_{xt},D_{xy},D_{xz}$\} can be selected to form a complete set of invariant
generators together with lower-degree invariants $H,L,M$. Because of the
matrix-determinant representation of these invariants, we can easily deduce
their permutational properties. For instance, if we take the expression for $%
L$ in Eq. (\ref{LMN}) and permute the first and third qubit, the elements in
its matrix transforms as follows: $a_{2}\leftrightarrow
a_{8},a_{3}\leftrightarrow a_{9},a_{6}\leftrightarrow
a_{12},a_{7}\leftrightarrow a_{13}$ and all other elements do not change. We
see immediately that this gives us $-N$ since the new matrix obtained after
these transformations is just the matrix of $N$ with transposition and
exchange of two columns. The permutational properties of other three
6-degree invariants can be obtained similarly. $H$ is obviously invariant
under qubit permutaitons as can be seen directly from Eq. (\ref{H}). These
results are summaried in Table I.

\begin{table}[h]
\begin{tabular}{c|c|c|c|c|c|c|c}
\hline\hline
qubit permutation & H & L & M & N & D$_{xt}$ & D$_{xy}$ & D$_{xz}$ \\ \hline
$1\leftrightarrow 2$ &  &  &  &  &  &  &  \\ \cline{1-1}
$3\leftrightarrow 4$ & \raisebox{2.3ex}[0pt]{H} & \raisebox{2.3ex}[0pt]{-L}
& \raisebox{2.3ex}[0pt]{-N} & \raisebox{2.3ex}[0pt]{-M} & %
\raisebox{2.3ex}[0pt]{D$_{xz}$} & \raisebox{2.3ex}[0pt]{D$_{xy}$} & %
\raisebox{2.3ex}[0pt]{D$_{xt}$} \\ \hline
$1\leftrightarrow 3$ &  &  &  &  &  &  &  \\ \cline{1-1}
$2\leftrightarrow 4$ & \raisebox{2.3ex}[0pt]{H} & \raisebox{2.3ex}[0pt]{-N}
& \raisebox{2.3ex}[0pt]{-M} & \raisebox{2.3ex}[0pt]{-L} & %
\raisebox{2.3ex}[0pt]{D$_{xy}$} & \raisebox{2.3ex}[0pt]{D$_{xt}$} & %
\raisebox{2.3ex}[0pt]{D$_{xz}$} \\ \hline
$1\leftrightarrow 4$ &  &  &  &  &  &  &  \\ \cline{1-1}
$2\leftrightarrow 3$ & \raisebox{2.3ex}[0pt]{H} & \raisebox{2.3ex}[0pt]{-M}
& \raisebox{2.3ex}[0pt]{-L} & \raisebox{2.3ex}[0pt]{-N} & %
\raisebox{2.3ex}[0pt]{D$_{xt}$} & \raisebox{2.3ex}[0pt]{D$_{xz}$} & %
\raisebox{2.3ex}[0pt]{D$_{xy}$} \\ \hline\hline
\end{tabular}%
\caption{Transformations of invariants under qubit permutations.}
\label{table1}
\end{table}

All these invariants, $\{L,M,N,D_{xt},D_{xy},D_{xz}\}$ can be also expressed
using anti-symmetric tensors. By directly expanding the determinant
expressions of $\{L,M,N\}$ and observing their terms, we obtain the
following relations: % between them and the invariants
%constructed with antisymmetric tensors:

\begin{widetext}
\begin{eqnarray}
4(N-M)
&=&A_{i_{1}j_{1}k_{1}l_{1}}A_{i_{2}j_{2}k_{2}l_{2}}A_{i_{3}j_{3}k_{3}l_{3}}A_{i_{4}j_{4}k_{4}l_{4}}
\label{N-M} \\
&&\times \epsilon _{i_{1}i_{2}}\epsilon _{i_{3}i_{4}}\epsilon
_{j_{1}j_{2}}\epsilon _{j_{3}j_{4}}\epsilon _{k_{1}k_{3}}\epsilon
_{k_{2}k_{4}}\epsilon _{l_{1}l_{4}}\epsilon _{l_{2}l_{3}},  \notag
\end{eqnarray}%
\begin{eqnarray}
4(M-L)
&=&A_{i_{1}j_{1}k_{1}l_{1}}A_{i_{2}j_{2}k_{2}l_{2}}A_{i_{3}j_{3}k_{3}l_{3}}A_{i_{4}j_{4}k_{4}l_{4}}
\label{M-L} \\
&&\times \epsilon _{i_{1}i_{2}}\epsilon _{i_{3}i_{4}}\epsilon
_{j_{1}j_{3}}\epsilon _{j_{2}j_{4}}\epsilon _{k_{1}k_{4}}\epsilon
_{k_{2}k_{3}}\epsilon _{l_{1}l_{2}}\epsilon _{l_{3}l_{4}},  \notag
\end{eqnarray}%
\begin{eqnarray}
4(L-N)
&=&A_{i_{1}j_{1}k_{1}l_{1}}A_{i_{2}j_{2}k_{2}l_{2}}A_{i_{3}j_{3}k_{3}l_{3}}A_{i_{4}j_{4}k_{4}l_{4}}
\label{L-N} \\
&&\times \epsilon _{i_{1}i_{2}}\epsilon _{i_{3}i_{4}}\epsilon
_{j_{1}j_{4}}\epsilon _{j_{2}j_{3}}\epsilon _{k_{1}k_{2}}\epsilon
_{k_{3}k_{4}}\epsilon _{l_{1}l_{3}}\epsilon _{l_{2}l_{4}}.  \notag
\end{eqnarray}%
\end{widetext}

By simple algebraic manipulations, we can express $L,M$ and $N$ in terms of
the antisymmetric tensor $\epsilon ,$ 
\begin{eqnarray}
L &=&\frac{1}{12}%
A_{i_{1}j_{1}k_{1}l_{1}}A_{i_{2}j_{2}k_{2}l_{2}}A_{i_{3}j_{3}k_{3}l_{3}}A_{i_{4}j_{4}k_{4}l_{4}}
\notag \\
&&\times \epsilon _{i_{1}i_{2}}\epsilon _{i_{3}i_{4}}(\epsilon
_{j_{1}j_{4}}\epsilon _{j_{2}j_{3}}\epsilon _{k_{1}k_{2}}\epsilon
_{k_{3}k_{4}}\epsilon _{l_{1}l_{3}}\epsilon _{l_{2}l_{4}}  \notag \\
&&-\epsilon _{j_{1}j_{3}}\epsilon _{j_{2}j_{4}}\epsilon
_{k_{1}k_{4}}\epsilon _{k_{2}k_{3}}\epsilon _{l_{1}l_{2}}\epsilon
_{l_{3}l_{4}}),  \label{AntiL} \\
M &=&\frac{1}{12}%
A_{i_{1}j_{1}k_{1}l_{1}}A_{i_{2}j_{2}k_{2}l_{2}}A_{i_{3}j_{3}k_{3}l_{3}}A_{i_{4}j_{4}k_{4}l_{4}}
\notag \\
&&\times \epsilon _{i_{1}i_{2}}\epsilon _{i_{3}i_{4}}(\epsilon
_{j_{1}j_{3}}\epsilon _{j_{2}j_{4}}\epsilon _{k_{1}k_{4}}\epsilon
_{k_{2}k_{3}}\epsilon _{l_{1}l_{2}}\epsilon _{l_{3}l_{4}}  \notag \\
&&-\epsilon _{j_{1}j_{2}}\epsilon _{j_{3}j_{4}}\epsilon
_{k_{1}k_{3}}\epsilon _{k_{2}k_{4}}\epsilon _{l_{1}l_{4}}\epsilon
_{l_{2}l_{3}}),  \label{AntiM} \\
N &=&\frac{1}{12}%
A_{i_{1}j_{1}k_{1}l_{1}}A_{i_{2}j_{2}k_{2}l_{2}}A_{i_{3}j_{3}k_{3}l_{3}}A_{i_{4}j_{4}k_{4}l_{4}}
\notag \\
&&\times \epsilon _{i_{1}i_{2}}\epsilon _{i_{3}i_{4}}(\epsilon
_{j_{1}j_{2}}\epsilon _{j_{3}j_{4}}\epsilon _{k_{1}k_{3}}\epsilon
_{k_{2}k_{4}}\epsilon _{l_{1}l_{4}}\epsilon _{l_{2}l_{3}}  \notag \\
&&-\epsilon _{j_{1}j_{4}}\epsilon _{j_{2}j_{3}}\epsilon
_{k_{1}k_{2}}\epsilon _{k_{3}k_{4}}\epsilon _{l_{1}l_{3}}\epsilon
_{l_{2}l_{4}}).  \label{AntiN}
\end{eqnarray}%
With the aid of a computer programme, we expand the complicated determinant
expressions for the degree-6 invariants and find that $D_{xz}+D_{xt}+D_{xy}$
can be simply expressed with the antisymmetric tensors as follows,%
\begin{eqnarray}
4(D_{xz}+D_{xt}+D_{xy})
&=&A_{i_{1}j_{1}k_{1}l_{1}}A_{i_{2}j_{2}k_{2}l_{2}}A_{i_{3}j_{3}k_{3}l_{3}}A_{i_{4}j_{4}k_{4}l_{4}}A_{i_{5}j_{5}k_{5}l_{5}}A_{i_{6}j_{6}k_{6}l_{6}}
\notag \\
&&\times \epsilon _{i_{1}i_{2}}\epsilon _{i_{3}i_{4}}\epsilon
_{i_{5}i_{6}}\epsilon _{j_{1}j_{3}}\epsilon _{j_{2}j_{4}}\epsilon
_{j_{5}j_{6}}\epsilon _{k_{1}k_{5}}\epsilon _{k_{2}k_{6}}\epsilon
_{k_{3}k_{4}}\epsilon _{l_{1}l_{2}}\epsilon _{l_{3}l_{5}}\epsilon
_{l_{4}l_{6}}.  \label{D}
\end{eqnarray}%
With the relations between $H,L,M,N$ and $D_{xt},D_{xy}$,$D_{xz}$, we get, 
\begin{eqnarray}
H(M-L) &=&2D_{xt}-D_{xy}-D_{xz}  \notag \\
&=&3D_{xt}-(D_{xz}+D_{xt}+D_{xy}).  \label{H(M-L)}
\end{eqnarray}%
The antisymmetric tensor expression of $D_{xt}$ is:

\begin{eqnarray}
D_{xt} &=&\frac{1}{3}[(D_{xz}+D_{xt}+D_{xy})+H(M-L)]  \notag \\
&=&\frac{1}{12}%
A_{i_{1}j_{1}k_{1}l_{1}}A_{i_{2}j_{2}k_{2}l_{2}}A_{i_{3}j_{3}k_{3}l_{3}}A_{i_{4}j_{4}k_{4}l_{4}}A_{i_{5}j_{5}k_{5}l_{5}}A_{i_{6}j_{6}k_{6}l_{6}}
\notag \\
&&\times \lbrack \epsilon _{i_{1}i_{2}}\epsilon _{i_{3}i_{4}}\epsilon
_{i_{5}i_{6}}\epsilon _{j_{1}j_{3}}\epsilon _{j_{2}j_{4}}\epsilon
_{j_{5}j_{6}}\epsilon _{k_{1}k_{5}}\epsilon _{k_{2}k_{6}}\epsilon
_{k_{3}k_{4}}\epsilon _{l_{1}l_{2}}\epsilon _{l_{3}l_{5}}\epsilon
_{l_{4}l_{6}}  \notag \\
&&+\frac{1}{2}\epsilon _{i_{1}i_{2}}\epsilon _{i_{3}i_{4}}\epsilon
_{i_{5}i_{6}}\epsilon _{j_{1}j_{2}}\epsilon _{j_{3}j_{5}}\epsilon
_{j_{4}j_{6}}\epsilon _{k_{1}k_{2}}\epsilon _{k_{3}k_{6}}\epsilon
_{k_{4}k_{5}}\epsilon _{l_{1}l_{2}}\epsilon _{l_{3}l_{4}}\epsilon
_{l_{5}l_{6}}].  \label{DXT}
\end{eqnarray}%
The antisymmetric tensor expressions of $D_{xy}$ and $D_{xz}$ can be
similarly obtained.

Now that we have antisymmetric tensor expressions for both $\left\vert 
\mathcal{F}_{i}\right\vert $'s and \{$H,L,M,D_{xt}$\}, we can find the
relations between them. By making many trials using a computer program, we
get 
\begin{equation}
\left\vert \mathcal{F}_{1}\right\vert =8[4(D_{xz}+D_{xt}+D_{xy})-H^{3}],
\label{monotone1}
\end{equation}%
\begin{eqnarray}
\left\vert \mathcal{F}_{2}\right\vert
&=&16[H^{4}-4H(2D_{xt}+D_{xz}+D_{xy})-16LM]  \notag \\
&=&16[H^{4}-4H(D_{xt}+D_{xz}+D_{xy})-4(HD_{xt}+4LM)],  \label{monotone2}
\end{eqnarray}%
\begin{eqnarray}
\left\vert \mathcal{F}_{3}\right\vert &=&32[4(N-M)+H^{2}]\times \lbrack
4(L-N)+H^{2}]\times \lbrack 4(M-L)+H^{2}]  \notag \\
&=&32\{H^{6}+16H^{2}[-(M^{2}+N^{2}+L^{2})  \notag \\
&&+(MN+NL+ML)]+64(N-M)(L-N)(M-L)\}.  \label{monotone3}
\end{eqnarray}%
These nontrivial identities are the key results that allow us to analyze the
permutational properties of $\left\vert \mathcal{F}_{i}\right\vert $'s. From
Table \ref{table1}, we see that $%
H,D_{xt}+D_{xz}+D_{xy},M^{2}+N^{2}+L^{2},(N-M)(L-N)(M-L)$ and $(MN+NL+ML)$
are all invariant under permutations of the four qubits. Therefore, $%
\left\vert \mathcal{F}_{1}\right\vert $ and $\left\vert \mathcal{F}%
_{3}\right\vert $ are both invariant under qubit permutations. However, $%
\left\vert \mathcal{F}_{2}\right\vert $ is not invariant under qubit
permutations because $(HD_{xt}+4LM)$ in $\left\vert \mathcal{F}%
_{2}\right\vert $ is not. This implies that $\left\vert \mathcal{F}%
_{2}\right\vert $ cannot be considered a good candidate for genuine 4-qubit
entanglement measures.

To address this difficulty, we propose a new permutation-invariant monotone
by using the following method. Starting with the expression for $\left\vert 
\mathcal{F}_{2}\right\vert $ and making permutations between qubits, we can
construct two other monotones according to Table \ref{table1}: 
\begin{eqnarray}
\left\vert \mathcal{F}_{4}\right\vert
&=&16[H^{4}-4H(D_{xt}+2D_{xz}+D_{xy})-16LN]  \notag \\
&=&16[H^{4}-4H(D_{xt}+D_{xz}+D_{xy})-4(HD_{xz}+4LN)],  \label{f4}
\end{eqnarray}%
\begin{eqnarray}
\left\vert \mathcal{F}_{5}\right\vert
&=&16[H^{4}-4H(D_{xt}+D_{xz}+2D_{xy})-16MN]  \notag \\
&=&16[H^{4}-4H(D_{xt}+D_{xz}+D_{xy})-4(HD_{xy}+4MN)].  \label{f5}
\end{eqnarray}%
Since the three monotones $\left\vert \mathcal{F}_{2}\right\vert ,\left\vert 
\mathcal{F}_{4}\right\vert ,\left\vert \mathcal{F}_{5}\right\vert $ are
obtained by qubit permutations, their sum $\left\vert \mathcal{F}%
_{2}^{\prime }\right\vert $ is then obviously permutation-invariant: 
\begin{equation}
\left\vert \mathcal{F}_{2}^{\prime }\right\vert
=16[3H^{4}-16H(D_{xt}+D_{xz}+D_{xy})-16(MN+NL+ML)].  \label{newf2}
\end{equation}

$\left\vert \mathcal{F}_{2}^{\prime }\right\vert $ is an entanglement
monotone because it is derived from $\left\vert \mathcal{F}_{2}\right\vert $%
, $\left\vert \mathcal{F}_{4}\right\vert $ and $\left\vert \mathcal{F}%
_{5}\right\vert $ which are themselves entanglement monotones. It yields
zero for all product states, and reaches 1 for a class of maximally
entangled states (see below). Together with the fact that it is
premutation-invariant, these suggest that $\left\vert \mathcal{F}%
_{2}^{\prime }\right\vert $ is potentially a good candidate for genuine
4-qubit entanglement measure. To see its advantage in characterizing genuine
multipartite entanglement, we calculate its value in comparison with $%
\left\vert \mathcal{F}_{2}\right\vert $ for three maximally entangled
4-qubit states 
\begin{eqnarray*}
\left\vert \Phi _{1}\right\rangle &=&\frac{1}{\sqrt{2}}(\left\vert
1111\right\rangle +\left\vert 0000\right\rangle ), \\
\left\vert \Phi _{2}\right\rangle &=&\frac{1}{\sqrt{6}}(\sqrt{2}\left\vert
1111\right\rangle +\left\vert 1000\right\rangle +\left\vert
0100\right\rangle +\left\vert 0010\right\rangle +\left\vert
0001\right\rangle ), \\
\left\vert \Phi _{3}\right\rangle &=&\frac{1}{2}(\left\vert
1111\right\rangle +\left\vert 1100\right\rangle +\left\vert
0010\right\rangle +\left\vert 0001\right\rangle ),
\end{eqnarray*}%
and states derived from them by qubit permutations. $\left\vert \Phi
_{1}\right\rangle ,\left\vert \Phi _{2}\right\rangle $ and $\left\vert \Phi
_{3}\right\rangle $ are known to be genuine 4-qubit entangled and belong to
different entanglement classes. As shown in the upper left (3 by 3) corner
of Table \ref{table2}, for $\left\vert \Phi _{1}\right\rangle ,\left\vert
\Phi _{2}\right\rangle $ and $\left\vert \Phi _{3}\right\rangle ,$ different
subsets of \{$\left\vert \mathcal{F}_{1}\right\vert ,\left\vert \mathcal{F}%
_{2}\right\vert ,\left\vert \mathcal{F}_{3}\right\vert $\} have zero values.
Since a state with zero value for one monotone can not be transformed into
another state with nonzero value for the same monotone under SLOCC
operations, $\left\vert \mathcal{F}_{i}\right\vert $'s are therefore a
powerful tool in distinguishing these three entanglement classes.

Now we consider the transformations of $\left\vert \Phi _{1}\right\rangle
,\left\vert \Phi _{2}\right\rangle ,\left\vert \Phi _{3}\right\rangle $
under qubit permutations. It is easy to see that states $\left\vert \Phi
_{1}\right\rangle ,\left\vert \Phi _{2}\right\rangle $ are both
permutation-invariant. However, we can obtain two other 4-qubit maximally
entangled states from $\left\vert \Phi _{3}\right\rangle $ by qubit
permutation: 
\begin{eqnarray*}
\left\vert \Phi _{4}\right\rangle &=&\frac{1}{2}(\left\vert
1111\right\rangle +\left\vert 1001\right\rangle +\left\vert
0010\right\rangle +\left\vert 0100\right\rangle ), \\
\left\vert \Phi _{5}\right\rangle &=&\frac{1}{2}(\left\vert
1111\right\rangle +\left\vert 0101\right\rangle +\left\vert
1000\right\rangle +\left\vert 0010\right\rangle ).
\end{eqnarray*}%
$\left\vert \Phi _{4}\right\rangle ,\left\vert \Phi _{5}\right\rangle $
should have the same genuine 4-qubit entanglement with $\left\vert \Phi
_{3}\right\rangle $ since they are obtained from $\left\vert \Phi
_{3}\right\rangle $ by qubit permutaitons. With local unitary operations, it
can be easily shown that they all belong to one of the two types of 4-qubit
graph states in \cite{Hein}. However, as shown in Table \ref{table2} $%
\left\vert \mathcal{F}_{2}\right\vert $ has different values for $\left\vert
\Phi _{3}\right\rangle ,\left\vert \Phi _{4}\right\rangle $ and these two
states would be characterized as belonging to different entanglement classes
should \{$\left\vert \mathcal{F}_{1}\right\vert ,\left\vert \mathcal{F}%
_{2}\right\vert ,\left\vert \mathcal{F}_{3}\right\vert $\} be used to study
the entanglement structures of 4-qubit pure states. In contrast, \{$%
\left\vert \mathcal{F}_{1}\right\vert ,\left\vert \mathcal{F}_{2}^{\prime
}\right\vert ,\left\vert \mathcal{F}_{3}\right\vert $\} yield the same
values for $\left\vert \Phi _{3}\right\rangle ,\left\vert \Phi
_{4}\right\rangle $ and $\left\vert \Phi _{5}\right\rangle $ and thus
correctly put them in the same entanglement class. This is a direct
manifestation of the permutation-invariance of $\left\vert \mathcal{F}%
_{2}^{\prime }\right\vert $.

\begin{table}[h]
\begin{tabular}{|c|c|c|c|c|c|c|}
\hline\hline
& $\left\vert \mathcal{F}_{1}\right\vert $ & $\left\vert \mathcal{F}%
_{2}\right\vert $ & $\left\vert \mathcal{F}_{3}\right\vert $ & $\left\vert 
\mathcal{F}_{4}\right\vert $ & $\left\vert \mathcal{F}_{5}\right\vert $ & $%
\left\vert \mathcal{F}_{2}^{\prime }\right\vert $ \\ \hline
$\left\vert \Phi _{1}\right\rangle $ & 1 & 1 & $\frac{1}{2}$ & 1 & 1 & 3 \\ 
\hline
$\left\vert \Phi _{2}\right\rangle $ & $\frac{8}{9}$ & 0 & 0 & 0 & 0 & 0 \\ 
\hline
$\left\vert \Phi _{3}\right\rangle $ & 0 & 0 & 1 & 0 & 1 & 1 \\ \hline
$\left\vert \Phi _{4}\right\rangle $ & 0 & 1 & 1 & 0 & 0 & 1 \\ \hline
$\left\vert \Phi _{5}\right\rangle $ & 0 & 0 & 1 & 1 & 0 & 1 \\ \hline\hline
\end{tabular}%
\caption{Entanglement monotones for the maximally entangled 4-qubit states.}
\label{table2}
\end{table}

To further see the value of $\left\vert \mathcal{F}_{2}^{^{\prime
}}\right\vert $, we calculate the genuine multipartite entanglement for
4-qubit cluster-class states as an interesting application of our results.
This problem has been considered in \cite{Bai1, Bai2}, where the authors
used the principle of quantum complementarity relations (QCR) to derive
genuine 4-qubit entanglement by subtracting fewer-partite entanglement from
a qubit's bipartite entanglement with the rest of the 4-qubit system. Though
QCR reveals some very intriguing physics, its procedure is rather
complicated and it is not easy to even prove that the result given by QCR is
a monotone \cite{Bai2}. It also relies on the availability of properly
defined fewer-partite entanglement measures \cite{Bai1}.

We re-examine the genuine 4-qubit entanglement in cluster-class states using 
$\left\vert \mathcal{F}_{2}^{^{\prime }}\right\vert $. In Ref.\cite{Bai2},
the authors considered three types of 4-qubit cluster-class states. These
states can be obtained by local operations starting from the maximally
entangled states $\left\vert \Phi _{1}\right\rangle $ and $\left\vert \Phi
_{3}\right\rangle $ ($\left\vert \Phi _{4}\right\rangle ,\left\vert \Phi
_{5}\right\rangle $). For example, the first type of states, 
\begin{equation*}
\left\vert \Pi _{1}\right\rangle =a\left\vert 0000\right\rangle +b\left\vert
0011\right\rangle +c\left\vert 1100\right\rangle -d\left\vert
1111\right\rangle ,
\end{equation*}%
correspond to our $\left\vert \Phi _{3}\right\rangle $ class. This is
because of the following transformations under local unitary operations: 
\begin{eqnarray}
\left\vert \Phi _{3}\right\rangle &=&\frac{1}{2}(\left\vert
1111\right\rangle +\left\vert 1100\right\rangle +\left\vert
0010\right\rangle +\left\vert 0001\right\rangle )_{1234}  \notag \\
&&\underrightarrow{\sigma _{4}^{x}}\frac{1}{2}(\left\vert 1110\right\rangle
+\left\vert 1101\right\rangle +\left\vert 0011\right\rangle +\left\vert
0000\right\rangle )_{1234}  \notag \\
&&\underrightarrow{H_{3}\otimes H_{4}}\frac{1}{2}(\left\vert
0000\right\rangle +\left\vert 0011\right\rangle +\left\vert
1100\right\rangle -\left\vert 1111\right\rangle )_{1234},  \label{equistate}
\end{eqnarray}%
\newline
where $H$ is the Hadamard transformation. Now we calculate the values of $%
\left\vert \mathcal{F}_{2}^{^{\prime }}\right\vert $ for $\left\vert \Pi
_{1}\right\rangle$. First, since $\left\vert \mathcal{F}_{i}\right\vert $'s
are all entanglement monotones, we have $\left\vert \mathcal{F}%
_{2}\right\vert =\left\vert \mathcal{F} _{4}\right\vert =0$ according to
Table II. This can also be deduced by the fact that $ad=bc$ since $%
\left\vert \Pi _{1}\right\rangle $ and $\left\vert \Phi _{3}\right\rangle $
can be converted into each other with finite probabilities through
invertible local operations. Therefore, $\left\vert \mathcal{F}%
_{2}^{^{\prime }}\right\vert =\left\vert \mathcal{F}_{5}\right\vert
=16|ad+bc|^{4}$. Since $\left\vert \mathcal{F}_{2}^{^{\prime }}\right\vert $
is of degree 8 and the result from QCR is of degree 4, we take the square
root of $\left\vert \mathcal{F}_{2}^{^{\prime }}\right\vert $ and we find
that the result is identical to the 4-qubit entanglement calculated from QCR
which is $4|ad+bc|^{2}$.

The second type of states that the authors considered in \cite{Bai2}, 
\begin{equation*}
\left\vert \Pi _{2}\right\rangle =a\left\vert 0000\right\rangle -b\left\vert
0111\right\rangle -c\left\vert 1010\right\rangle +d\left\vert
1101\right\rangle
\end{equation*}%
correspond to the $|\Phi _{5}\rangle $ class in Table II since they can be
obtained from $|\Phi _{5}\rangle $ by local opearions similar to Eq.(\ref%
{equistate}). Interestingly, if we calculate the 4-qubit entanglement of $%
|\Pi _{2}\rangle $ by QCR we will need to calculate 3-qubit entanglement
first, therefore the calculation is nontrivial. When we calculate $%
\left\vert \mathcal{F}_{2}^{^{\prime }}\right\vert $, we find $\left\vert 
\mathcal{F}_{2}\right\vert =\left\vert \mathcal{F}_{5}\right\vert =0$ and $%
\left\vert \mathcal{F}_{2}^{^{\prime }}\right\vert =\left\vert \mathcal{F}%
_{4}\right\vert =256|abcd|^{2}$. This is exactly the square of the result
obtained from QCR.

Therefore, the 4-qubit genuine entanglement of cluster-class states in \cite%
{Bai2} calculated according to QCR coincides with the value obtained with $%
\left\vert \mathcal{F}_{2}^{^{\prime }}\right\vert $. Considering that these
two approaches have no obvious intrinsic connections, this is intriguing and
it hints that $\left\vert \mathcal{F}_{2}^{^{\prime }}\right\vert $ may
indeed be a good genuine 4-qubit entanglement measure. We can also
conveniently draw some conclusions not so easy to prove in QCR. E.g., the
result given by QCR is a monotone and permutation-invariant. The advantage
of $\left\vert \mathcal{F}_{2}^{^{\prime }}\right\vert $ though, is it can
be calculated directly once a state is given without reference to
fewer-partite entanglement. Therefore it may find broader applications,
especially when the application of QCR is difficult due to the difficulty in
determining what fewer-partite entanglement expressions should be used \cite%
{Bai1}.

In conclusion, we studied the permutational properties of multipartite
entanglement monotones by specifically examining the three 4-qubit
entanglement monotones \{$\left\vert \mathcal{F}_{1}\right\vert ,\left\vert 
\mathcal{F}_{2}\right\vert ,\left\vert \mathcal{F}_{3}\right\vert \}$
introduced by Osterloh and Siewert\cite{Osterloh,Osterloh1}. We find one of
these, $\left\vert \mathcal{F}_{2}\right\vert$, does not satisfy the natural
requirement of permutational invariance for a genuine multipartite
entanglement measure and propose an alternative that is
permutation-invariant. By comparison with results from QCR we find that our
new monotone properly measures the genuine 4-qubit entanglement in 4-qubit
cluster-class states. Our results are intriguing in understanding
multipartite entanglement.

\textit{Note added: After we finished the preparation of our manuscript, we
noticed that the relations in Eq.(\ref{monotone1},\ref{monotone2},\ref%
{monotone3}) were also obtained by D. \v{Z}. okovi\'{c} and A. Osterloh in a
recent work \cite{Dokovic}. }

\begin{acknowledgments}
This work was funded by National Fundamental Research Program 2006CB921900,
NCET-04-0587, the Innovation funds from Chinese Academy of Sciences, and
National Natural Science Foundation of China (Grant No. 60621064, 10574126).
\end{acknowledgments}

\end{document}